\documentclass[aip,sd,amsmath,amssymb,floatfix]{revtex4-1}
\pdfoutput=1
\usepackage{mathtools}
\usepackage{graphicx}
\usepackage{amssymb}
\usepackage{amsmath}
\usepackage{epstopdf}
\usepackage{diagbox}
\usepackage{bbm}
\usepackage{dsfont}
\usepackage{bbold}
\usepackage{color,soul}
\usepackage{dcolumn}
\usepackage{soul}
\usepackage{latexsym}
\usepackage{bm}
\usepackage{upgreek}
\usepackage{ulem}
\usepackage{listings}
\usepackage{cases}
\newcommand{\xj} {
J
}
  
\begin{document}

\title{Renormalized site density functional theory}

\author{Gennady N. Chuev}
\thanks{genchuev@rambler.ru}
\affiliation{
 Institute of Theoretical and Experimental Biophysics, Russian
 Academy of Science, Pushchino, Moscow Region, 142290, Russia}

\author{Marina V.~Fedotova}
\affiliation{G.A. Krestov Institute of Solution Chemistry, Russian Academy of Sciences, Akademicheskaya st., 1, 153045
Ivanovo, Russia}

\author{Marat Valiev}
\thanks{marat.valiev@pnl.gov}
\affiliation{Molecular Sciences Software Group, Environmental Molecular Sciences Laboratory,Pacific Northwest National Laboratory, WA 99352 USA}

\date{\today}

\begin{abstract}
Site density functional theory (SDFT) provides a rigorous framework for statistical mechanics analysis of inhomogeneous molecular liquids. The key defining feature of these systems is the presence of two very distinct interactions scales (intra- and inter-molecular), and as such proper description of both effects is critical to the accuracy of the calculations. Current SDFT applications utilize the same approximation scheme for both interaction motifs, which negatively impacts the results.
Dual space methodology, used in this work, alleviates this issue by providing the flexibility of evaluating part of the interactions in traditional field representation. For molecular liquid this translates into retaining density representation for inter-molecular interactions but describing stiff intra-molecular remainder by more appropriate conventional field based methods.
This opens the way to decouple analysis of the two interactions scales - the idea which is developed further in this work for the case of homogeneous reference approximation of inter-molecular interactions. We demonstrate that by defining new collective variables, the behaviour of the original molecular liquid system at the inter-molecular level can be transformed to resemble that of an effective simple fluid mixture. The latter is linked to intra-molecular scale through renormalized interaction parameters. We illustrate this renormalization procedure for several types of diatomic liquids, showing that this approach cures many of the shortcomings of existing SDFT methods. 

\end{abstract}

\maketitle

\section{Introduction}

Site density-functional theory (SDFT) provides statistical mechanics description of 
\textit{molecular} many-body systems  in terms of the density averages of atomic sites in the molecule.
The concept is similar to electronic structure DFT, but applied to nuclei degrees of 
freedom, treated classically.  From practical point of view, the main advantage of 
SDFT is that it provides direct approximation of phase space averages, 
thus avoiding numerical expense of sampling with 
molecular dynamics or Monte-Carlo methods.

Despite the classical nature of the problem, development of 
practically applicable SDFT approaches for molecular systems 
has proven to be challenging. While the problem has been analyzed some thirty years ago by  
Chandler, McCoy, and Singer 
in what is known as a CMS approach,\cite{Chandler1986a,Chandler1986b}
applying these ideas 
in practice has not been easy. 
The primary reason is the multi-scale nature of interactions, comprised of strong localized interactions 
that bind atoms into molecules (intra-molecular) and 
weak long range interactions between molecular units (inter-molecular). 
Given the distinct disparity of the two scales, they cannot be properly treated within the same approximation scheme, which presents significant theoretical 
challenges.

Currently most SDFT applications are based on 3D-RISM\cite{Beglov1997} approach, which has been applied quite successfully to a wide range of chemical
and biological  molecular systems \cite{ Chuev2006,Imai2007,Ikuta2007,Chuev2009,Howard2010,
Palmer2011,Fedotova2014,Yesudas2015,Fedotova2015,Terekhova2015,Phanich2016,Fedotova2017a,Roy2017,Maruyama2017,Fedotova2017,Hasegawa2017,
Hay2018,Fedotova2019,Cao2019,Fedotova2020,Fedotova2020a}. The key assumption in the method is that correlation effects in molecular liquids can be represented by second order density expansion around homogeneous reference system.
While this approximation works well for inter-molecular interactions, it is poorly suited for description of stiff intra-molecular potentials (aka chemical bonds).\cite{Chuev2020}
Recognizing this issue, we have developed an approach\cite{Valiev2018} that from the outset separated the treatment of intra- and inter-molecular interactions, employing an implicit scheme for the construction of density functional, akin to Kohn-Sham scheme in electronic structure DFT. The approach has shown promising results in curing the existing artifacts of RISM methods.\cite{Chuev2020}

The present work extends our previous investigation in two important ways. We show that our implicit scheme of construction density functional can be reformulated as a mixed or dual space formulation. The latter provides the flexibility to apply density-based representation only to part of the overall interactions, and retain conventional field-based treatment for the rest. Taking this approach in conjunction with homogeneous density approximation for inter-molecular interactions, we then demonstrate how our original molecular liquid can be transformed to resemble the effective simple fluid mixture. The overall process can be thought of as a renormalization procedure, such that the details of interaction at the intra-molecular scale are propagated to long range inter-molecular scale via effective interaction parameters. Applications to several representative types of diatomic liquids show that resulting approach provides much better accuracy than existing RISM methods, yet retaining computational simplicity of the latter.

The paper is organized as follows. We start by describing dual space formulation for molecular liquid systems and resulting system of self-consistent equations (Sections IIA and IIB). The renormalization procedure along with corresponding computational implementation is described in Section IIC and IID. Finally, in Section III we provide illustrative applications to various types of diatomic liquids.

\section{Methodology}
\subsection{Dual space representation}

Let us consider inhomogeneous molecular liquid consisting of a single molecule type with $M$ distinct atom sites. We assume that the interactions between atomic sites ($U$) can be separated into intra-molecular ($U_m$) and inter-molecular ($U_l$) contributions. The system is also subject to an external site dependent potential $v_\alpha(\mathbf{r}_{i\alpha})$, where 
$\mathbf{r}_{i\alpha}$ denotes coordinates of site $\alpha$ in molecule $i$. 

The natural starting point for the statistical mechanics analysis of the system is given by the grand thermodynamic potential or, as we will refer to it, a field functional:
\begin{equation}
\label{W[j]}
\Omega[\bm{\xj}] =  - \frac{1}{\beta }\ln \mathbb{Tr} \: e^{
- \beta
( 
U(\{\mathbf{r}_{i\alpha}\})
+
\sum v_\alpha(\mathbf{r}_{i\alpha})  
+
\sum \xj_\alpha(\mathbf{r}_{i\alpha})  
)
}
\end{equation}
where  $\mathbb{Tr}$ denotes grand canonical average
\begin{equation}
\mathbb{Tr}\left[  \ldots  \right] = \sum\limits_{N = 0}^\infty 
 \frac{e^{\beta\mu N}}{\Lambda^{3N} N!}\int {\left[  \ldots  \right]} \prod\limits_{i\alpha}  d{{\bf{r}}_{i\alpha}}
\end{equation}
The field functional is defined of over the space of auxiliary fields $\bm{\xj}(\bm{r})\equiv\{\xj_{\alpha}(\bm{r})\}$, with original system located at the point where $\bm{\xj}(\bm{r})=0$. 
One may say that it provides \textit{field-based representation} of the system.

Alternatively, the system can be also analyzed in the density domain - the premise behind the SDFT approach.\cite{Chandler1986a} The transformation from field to density based view is facilitated by the fact that the two variables form a conjugate pair
\begin{equation}
\label{dW/dJ}
\frac{{\delta \Omega[\bm{\xj}]}}{{\delta {\xj_\alpha}({\bf{r}})}} =  {\rho _\alpha}({\bf{r}}),
\end{equation}
which, via the Legendre transform, leads to the definition of the density functional $\Gamma[\bm{\rho}]$
\begin{equation}
\label{gamma_def}
\Gamma[\bm{\rho}]=\Omega[\bm{\xj}]-\bm{\xj}\cdot\bm{\rho}
\end{equation}
Here $\bm{\xj}$ itself is considered to be a functional of $\bm{\rho}$ by virtue of (\ref{dW/dJ}), with dot product representing integration and summation over site coordinates and indices.
\begin{eqnarray}
\boldsymbol{J}\cdot\boldsymbol{\rho} &\equiv& \sum_{_{\alpha}}\int {J}_{\alpha}(\mathbf{r}) \rho_\alpha(\mathbf{r})d\mathbf{r}
 \label{product}
\end{eqnarray}

One of the main advantages of working in density space is that solution can now be posed as a variational problem
\begin{equation}
\label{var}
\frac{\delta \Gamma[\bm{\rho}]}{\delta \bm{\rho}({\bf{r}})}=0
\end{equation}
Density-based presentation has been used widely in the theory of simple liquids. \cite{Evans1979}
However, its application to systems such as molecular fluids has been challenging. The main complication is that unlike generating functional the functional form of the density functional is generally unknown. 

The approach that we use in this work, builds up upon ideas presented in our earlier
investigation\cite{Valiev2018} and looks at the molecular liquid problem from the standpoint of mixed field/density representation. The main idea behind this dual space formulation, is to enable the use of different domains, field or density, when evaluating contributions from different parts of the interaction potential. In the case of molecular liquid, the intra-molecular correlations can be easily analyzed in field domain, but exceedingly difficult to treat in density representation.  On the other hand, overwhelming success of density based methods in  simple liquids, indicates that density domain is a natural setting for the treatment of long range inter-molecular interactions. One of the mechanisms, by which we can formally separate the two contributions consists in defining an intermediate, molecular gas system. By construction, the latter contains only intra-molecular contributions ($U=U_m$) that can be analyzed in field domain using by means of the field functional
\begin{align}
    \Omega_m[\bm{\xj}] = \Omega[\bm{\xj}]\big\vert_{U=U_m}
    \label{Wm_general}
\end{align}
At the same time the inter-molecular interactions in density space can be identified as
\begin{align}
   \Upsilon_l[\bm{\rho}] = \Gamma[\bm{\rho}] - \Gamma_m[\bm{\rho}] 
   \label{upsilon}
\end{align}
where $\Gamma_m[\bm{\rho}]= \Gamma[\bm{\rho}]\big\vert_{U=U_m}$ refers to the same molecular gas system.
The two contributions can be combined by defining the following dual space functional (see Appendix \ref{app:sec_M})
\begin{align}
    \mathcal{M}[\bm{\xj},\bm{\rho}]=\Omega_{m}\left[\bm{\xj}\right]+\Upsilon_l[\bm{\rho}] 
    -\bm{\xj} \cdot \bm{\rho}
    \label{M-functional}
\end{align}
It should be understood that in the definition of dual space functional, $\bm{\xj}$ and
$\bm{\rho}$ are playing role of \textit{independent variables}. The solution to our molecular liquid system is obtained at the extremum of $\mathcal{M}[\bm{\xj},\bm{\rho}]$ 
\begin{equation}
\frac{\delta \mathcal{M}[\bm{\xj},\bm{\rho}]}{\delta \bm{\xj}({\bf{r}})}=0
\label{M_extremum-J}
\end{equation}
\begin{equation}
\frac{\delta \mathcal{M}[\bm{\xj},\bm{\rho}]}{\delta \bm{\rho}({\bf{r}})}=0
\label{M_extremum_rho}
\end{equation}
In particular one can explicitly verify that the above conditions are completely equivalent to the implicit SDFT procedure presented earlier.\cite{Valiev2018}

For the purposes of this work, we will take one extra step and explicitly separate 
external potential $\bm{v}(\mathbf{r})$  by shifting the auxiliary potential as 
$\bm{\xj}(\mathbf{r}) \rightarrow \bm{\xj}(\mathbf{r}) - \bm{v}(\mathbf{r})$.
As a result of this transformation the intra-molecular contribution can be expressed simply as
\begin{equation}
\label{Wm[j]}
\Omega_m[\bm{\xj}] =  - \frac{1}{\beta }\ln \mathbb{Tr} \: e^{
- \beta
( 
U_m(\{\mathbf{r}_{i\alpha}\})
+
\sum \xj_\alpha(\mathbf{r}_{i\alpha})  
)
}
\end{equation}
and we acquire an additional external potential contribution in the dual space functional 
\begin{align}
    \mathcal{M}[\bm{\xj},\bm{\rho}]=\Omega_{m}\left[\bm{\xj}\right]+\Upsilon_l[\bm{\rho}] 
    -\bm{\xj} \cdot \bm{\rho}  + \bm{v} \cdot \bm{\rho}
    \label{M-functional-1}
\end{align}

\subsection{Self-consistent equations} 

The ability to separate the treatment of intra- and inter-molecular contributions into their respective, field and density, domains significantly simplifies the analysis of the molecular liquid problem and results in the self-consistent procedure akin to electronic structure DFT. We will illustrate this in the context of cluster-based expression\cite{Valiev2018} of molecular gas field functional $\Omega_m[\bm{\xj}]$
\begin{equation}
\Omega_m[\bm{\xj}]  = 
\Omega_m^{id}[\bm{\xj}] 
+
\Delta\Omega_m^{}[\bm{\xj}] 
\label{Wm_cluster}
\end{equation}
Here the 1st term represents an ideal gas contribution
\begin{align}
    \Omega_m^{id}[\bm{\xj}] =  -\frac{\rho_0}{\beta}
    \left( 1 +
    \sum_{\alpha=1}^M \int f_{\alpha}({\bf{r}}) d{\bf r} 
    \right)
\end{align}
where 
$\bm{f}$ represent the so-called Mayer functions
\begin{align}
\label{f}
  \bm{f}(\mathbf{r})=
        e^{
    -\beta 
       \bm{\xj}(\mathbf{r})
    }
    -1    
\end{align}
The second term contains remaining correlation contributions
\begin{equation}
\Delta\Omega_m[\bm{\xj}]  = 
-\frac{\rho_0}{\beta}
 \sum_{s=2}^{M}
   \frac{\mathtt{Tr}[\bm{D}^{(s)}\bm{f}^{s}] }{s !}
   \label{Wm_cluster}
\end{equation}
where $\bm{D}^{(s)}$ denote intra-molecular correlation functions.

The density of the molecular gas system can be obtained as a functional derivative of $\Omega_m[\bm{\xj}]$ with respect to $\bm{\xj}$ (cf. (\ref{dW/dJ}))
\begin{align}
     \bm{\rho}_m([\bm{\xj}],{\bf{r}})=
          \rho_0
     \left (
     1+\bm{\xi}([\bm{\xj}],{\bf{r}})
     \right)
 e^{-\beta 
\bm{J}({\bf{r}})
}
\label{rho_m}
\end{align}
where we defined a correlation hole functional as
\begin{align}
\label{xi}
    \bm{\xi}([\bm{\xj}],{\bf{r}}) = -\frac{\beta}{\rho_0}\frac{\delta \Delta\Omega_m[\bm{\xj}]}{\delta \bm{f}({\bf{r}})} 
    = \sum_{s=2}^{M}
   \frac{\mathtt{Tr}[\mathbb{D}^{(s)}\bm{f}^{s-1}] }{(s-1) !}
\end{align}

%[MORE DISCUSSION CAN GO HERE]

With the above results at hand, let us now consider the first extremum condition for $\mathcal{M}[\bm{\xj},\bm{\rho}]$, see Eq.(\ref{M_extremum-J}).
Performing variation with respect to $\bm{\xj}$ we immediately obtain the following important relationship
\begin{align}
     \boldsymbol{\rho}({\bf r})=
 \bm{\rho}_m([\bm{\xj}],{\bf{r}})
\label{rho_condition}
\end{align}
The above result essentially states that density of our molecular liquid system $\bm{\rho}({\bf r})$ can be represented as the density of the molecular gas system $\bm{\rho}_m({\bf r})$ at some (yet unknown) external field $\bm{\xj}({\bf r})$.
The situation in that sense is very similar to the Kohn-Sham procedure in electronic structure DFT. Of course, Eq. (\ref{rho_condition}) only provides half of the solution, we still do not know at which  $\bm{\xj}({\bf r})$ the equality (\ref{rho_condition}) takes place. The answer to that question is provided by the second variational condition (\ref{M_extremum_rho}). Application of the latter shows that the required field $\bm{\xj}({\bf r})$ consists of two components
\begin{align}
    \bm{J}({\bf{r}}) =  \bm{\upsilon}({\bf{r}}) + \bm{\phi}_l([\bm{\rho}],{\bf{r}})
    \label{scf_potential}
\end{align}
First one is the original external potential experienced by the system, and second is the inter-molecular correlation potential
\begin{align}
    \bm{\phi}_l([\bm{\rho}],{\bf{r}}) = \frac{\delta \Upsilon_l[\bm{\rho}]}{\delta \bm{\rho}({\bf{r}})}
    \label{phi_l}
\end{align}
The inter-molecular correlation potential can be viewed as an analog of exchange-correlation potential in Kohn-Sham density functional theory. It depends on the density and together with Eq. (\ref{rho_condition}) sets up a self-consistent system of equations
\begin{numcases}
\bm{\rho}({\bf r})=
          \rho_0
     \left (
     1+\bm{\xi}([\bm{\xj}],{\bf r})
     \right)
 e^{-\beta 
\bm{J}({\bf{r}})
}
\label{rho_scf}
\\
    \bm{J}({\bf{r}}) =  \bm{\upsilon}({\bf{r}}) + \bm{\phi}_l([\bm{\rho}],{\bf{r}})
\label{J_scf}
\end{numcases}
The remaining issue of evaluation of inter-molecular correlation will be discussed in the next section. 

\subsection{Renormalization procedure for inter-molecular interactions}
\label{sec:renormalization}
Evaluation of inter-molecular contributions on their own, in density domain, proves instrumental in building connections between molecular liquid and simple liquid problems. One particular technique, commonly used in the theory simple liquids, is the 2nd order expansion of excess free energy functional around homogeneous system, also known as HNC approximation. The excess free energy functional maps directly onto the inter-molecular correlation functional,
and it seems reasonable to employ the same approach in our case as well. Expanding the inter-molecular correlation functional 
in terms of density fluctuation,
$\Delta\bm{\rho}(\mathbf{r}) = \bm{\rho}(\mathbf{r})-\rho_0$,
 and keeping only the 2nd order terms we obtain
\begin{align}
\Upsilon_l[\bm{\rho}]  = \Upsilon_l[\rho_0] +\frac{1}{2\beta\rho_0}
\iint
\Delta\bm{\rho}(\mathbf{r})
\left[
\mathbf{S}^{-1}(|\mathbf{r}-\mathbf{r}'|) - \mathbf{S}_m^{-1}(|\mathbf{r}-\mathbf{r}'|)
\right]
\Delta\bm{\rho}(\mathbf{r}')
 d\mathbf{r}d\mathbf{r}'
\label{HNC}
\end{align}
Here  $\mathbf{S}$ and  $\mathbf{S}_m$  stand for structure factors for molecular liquid and molecular gas homogeneous systems respectively:
\begin{equation}
    \mathbf{S}_m = \bm{1} + \mathbf{D}^{(2)},
    \qquad
    \mathbf{S} = \mathbf{S}_m + \rho_0\mathbf{H}
\end{equation}
where $\mathbf{H}$ inter-molecular correlation
function, whose elements can be obtained from site-site radial distribution functions as $H_{\alpha\alpha'}(r)=g_{\alpha\alpha'}(r)-1$. 

The corresponding expression for the inter-molecular correlation potential can be obtained
from (\ref{phi_l}) and written in reciprocal space as
\begin{align}
\bm{\phi}_l(\mathbf{k}) = 
\frac{1}{\beta\rho_0}
      \left[
\mathbf{S}^{-1}(k)
-
\mathbf{S}_m^{-1}(k)
      \right ]
      \Delta\bm{\rho}(\mathbf{k})
      \label{phi_hnc_kspace}
\end{align}
Formally speaking, the above equations provide a perfectly valid extension of HNC approximation to molecular liquids. Yet, in practical applications, evaluation of inter-molecular potential as given by (\ref{phi_hnc_kspace}), will encounter significant numerical problems. The issue lies in in molecular structure factor $\mathbf{S}_m$, which becomes ill-conditioned for low $k$-values and completely degenerate for $k=0$:
\begin{align}
    \lim_{k\rightarrow 0} [\mathbf{S}_m(k)]_{\alpha\beta} = 1,
    \qquad \forall \alpha,\beta
\end{align}
While this issue may at first appear as a mere numerical inconvenience of HNC approximation, the roots of the problem run much deeper than that and reflect inherent multi-scale nature of the molecular liquid problem. Indeed, one of the major tenets of site-density description is that atomic site densities are treated as independent first class entities. This view is indeed very much appropriate and arguably necessary at the microscopic level. Yet from macroscopic perspective, our system is made from molecules, and it is precisely this discord between the two views that is being manifested here. To put in this on quantitative grounds, we observe that by virtue of being part of the same molecule, the atomic site densities in the limit of $k\rightarrow 0$, will start approaching the same value and eventually become degenerate 
\begin{align}
    \lim _{k \rightarrow 0}\rho_{\alpha}(\mathbf{k}) = \int \rho_{\alpha}(\mathbf{r}) d\mathbf{r} = N,  \qquad \forall \alpha
    \label{rho_limit}
\end{align}
At the same time, the density equation (see (\ref{rho_scf})) in this limit reduces to simple linear response expression
 \begin{align}
    \Delta\bm{\rho}(\mathbf{k}) \approx
    - \beta\rho_0\mathbf{S}_m(k)\bm{J}(\mathbf{k})
    \label{rho_response}
\end{align}
Thus the only way to satisfy degeneracy condition (\ref{rho_limit}) in general case, is for the structure factor $\mathbf{S}_m$ itself to become degenerate.
In other words, the degeneracy of molecular structure factor plays an essential role in ensuring the proper transition between microscopic and macroscopic scales for the molecular liquid problem.

To address the problem highlighted above we follow the common strategy of reformulating the problem in terms of some new renormalized variables. The intent is to transform  the original "bare" system into an effective one, such that the divergent terms can be compartmentalized into new interaction parameters. The simplest effective model for the molecular liquid system is that based on the simple fluid mixture, where each atomic site maps into a different component. The transformation is accomplished by defining the following variable
\begin{align}
     \bm{\bar{c}}(\mathbf{k}) = \frac{1}{\rho_0}\mathbf{S}_m(k)\mathbf{S}^{-1}(k)\Delta\bm{\rho}(\mathbf{k}) 
     \label{c_bar}
\end{align}
We can observe that in the absence of intra-molecular correlations 
($\mathbf{S}_m\rightarrow \mathbf{1}$), it reduces to a familiar direct correlation
function from simple liquid theory, $ \bm{c} = \mathbf{S}^{-1}\Delta\bm{\rho}/\rho_0 $.\textbf{\cite{Henderson1992}} In that sense, $\bm{\bar{c}}$ may be referred to as renormalized direct correlation function, however, in the context of this work, we choose to call it renormalized response function. Formulated in terms of $\bm{\bar{c}}$, the expression for inter-molecular correlation potential becomes
\begin{align}
     \beta\bm{\phi}_l({\bf{k}})= -
\mathbf{\bar{H}}(\mathrm {k})
      \bm{\bar{c}}({\bf{k}})
      \label{phi_renorm}
\end{align}
The above indeed identical in structure to simple fluid mixture expression, with interaction mediated by renormalized inter-molecular correlation functions (RCF) 
\begin{align}
      \bf{\bar{H}}(\mathrm {k})& \equiv 
      \mathbf{S}_m^{-1}(\mathrm {k} )\mathbf{H}(\mathrm {k})\mathbf{S}_m^{-1}(\mathrm {k} )
      \label{H_bar}
\end{align}
We should note that similar functions were previously obtained as a result of partial wave expansion of the molecular Ornstein–Zernike equation.\cite{Ten-no1999}

The density or the closure equation in terms of new variables takes the following form
\begin{equation}
\bm{\bar c}({\bf r})=
     \left [
     1+\bm{\xi}(\bm{\phi}_l,{\bf r})
     \right]
 e^{-\beta (\bm{\upsilon}({\bf{r}})+\bm{\phi}_l({\bf{r}}))}+\beta [\bm{S}_m \ast\bm{\phi}_l]({\bf{r}}) - 1
\label{closure_renorm}
\end{equation} 

Relations (\ref{closure_renorm}) and (\ref{phi_renorm}) form a complete  set of self-consistent equations for the analysis of inhomogeneous molecular 
liquids within homogeneous approximations. In particular, we note that low-$k$ values the solution can be obtained trivially as $
     \bm{\bar c}(\mathbf{k}) =  -\beta\mathbf{S}_m\bm{\upsilon}(\mathbf{k})$.
This implies that similar to site densities, renormalized response functions asymptotically at large distances also become degenerate and tending to the same value: $
     \lim_{r\rightarrow \infty} {\bar c}_{\alpha}(\mathbf{r}) =  -\beta\sum_{\alpha'=1}^{M} \upsilon_{\alpha'}(\mathbf{r})$.
The above should be contrasted with similar behaviour of direct correlation function in simple liquids: $
     \lim_{r\rightarrow \infty} c(\mathbf{r}) =  -\beta \upsilon(\mathbf{r})$.
     
The renormalization procedure presented above shifts the numerical problem of ill-conditioned structure factors to a more physical one - construction of proper effective correlation functions, i.e. RCF's. The latter still exhibit divergences for
low-$k$ values, which can be dealt it one of two ways. The first is based on regularization of their low-$k$ behaviour. Second, arguably more fundamental, way involves RCF's directly from MD simulations or approximately from liquid state theories. These issues will be discussed in more details on example applications to diatomic molecular liquids.

\subsection{Computational procedures}

The self-consistent procedure for solving renormalized site-density theory (RSDFT) equations (\ref{phi_renorm}) and (\ref{closure_renorm})  is quite similar to that in 3D-RISM\cite{Beglov1997}, employing Fast Fourier Transformation (FFT) for calculation of convolution type integrals in reciprocal space. The two major differences are the additional calculations related to correlation hole $\bm{\xi}$ and renormalized correlation function $\mathbf{\bar{H}}$, see Eqns. (\ref{xi}) and  (\ref{H_bar}). 

As we discussed in our previous work\cite{Chuev2020} the correlation hole plays a critical role in capturing chemical bond effects. For diatomic liquids considered here, it takes particular simple form as
\begin{equation}
\xi_{\alpha}({\bf k})= \sum_{\alpha'}(1-\delta_{\alpha\alpha'}) d(k)  f_{\alpha'}({\bf k}) 
\end{equation}
where $d(k)$ is the intra-molecular pair-correlation function 
\begin{align}
    d(k) = \frac{\sin(kl)}{kl}
    \label{d-function}
\end{align}
and $l$ is the bond length. As suggested by the above expression,  during calculations, $\bf{\xi}$ is first assembled in $k$-space and then transformed into real space using Fast Fourier Transform (FFT). 

Evaluation of renormalized correlation function (\ref{H_bar}) involves several steps. First we determine bare correlation functions from radial distribution function obtained from molecular dynamics (MD) simulations of homogeneous system, $H_{\alpha\beta}(r) = g_{\alpha\beta}(r) - 1$.
Due to  finite size of the simulation box and finite duration of MD simulations, the resulting correlation functions is noisy, and we smooth it out using the procedure described in the previous work.\cite{Chuev2013,Chuev2014} In the next step we assemble $\mathbf{\bar{H}}$ from (\ref{H_bar}) using analytical expression of inverse of intra-molecular structure factor. As discussed previously, the latter
will diverge at $k=0$ as 
$
    \mathbf{S}_m^{-1}(k) \sim O\left( k^{-2} \right)$,
To regularize the corresponding divergence in
$\mathbf{\bar{H}}(k) \sim O\left( k^{-2} \right)$, we use a simple modification
\begin{equation}
\mathbf{\bar{H}}(k) \rightarrow \mathbf{\bar{H}}(k) 
\frac{k^2}{k^2+k_0^2}
\label{h_bar_renorm}
\end{equation}
where $k_0$ is a cutoff parameter.
As a result of this modification the behaviour around $k=0$ becomes
$ \mathbf{\bar{H}}(k) \sim O\left( (k^2+k_0^2)^{-1} \right)$. The cutoff parameter $k_0$ is chosen such that short-ranged behaviour of site densities is not affected, i.e.   $k_0\sigma <<1$, where
$\sigma$ is a characteristic scale of oscillations in site densities. Our  
numerical tests have indicated that  $k_0 \le 0.03 $\AA$^{-1}$ does not affect the quality of 
the calculations for the systems under the consideration.

Aside additional calculations of $\bf{\xi}$ and $\bf{\bar{H}}$, the overall computational procedure is very similar to that used in 3D-RISM,
involving iterative solution with the use of FFT and the DIIS methods \cite{Kovalenko1999}.  
All the calculations have been performed with the use of NWChem package \cite{Apra2020} which involves the 
1D-RISM code \cite{Chuev2012}.
We have utilized equidistant real space grid consisting of $2^{16}$ points with the grid step equal to $0.02$\AA. 

Based on the initial value for $\bm{\phi}_l({\bf{r}})=0$, we evaluate 
self-consistent potential $\bm{J}(\bm{r})$ (\ref{J_scf}) and Mayer function $\bm{f}(\bm{r})$ (\ref{f}). Next, using FFT we transform Mayer function into reciprocal space and calculate the correlation hole $\bf{\xi}(\bm{k})$ (\ref{xi}). Next, we evaluate the renormalized response functions  by  (\ref{closure_renorm}). Using FFT we obtain $\bm{\bar{c}}({\bf{k}})$ and calculate 
new value of intermolecular correlation potential   $\bm{\phi}_l({\bf{k}})$ by (\ref{phi_renorm}). Finally  we obtain a new guess for  $\bm{\phi}_l({\bf{r}})$, making the inverse FFT.
 The iteration procedure is stopped when the relative changes in the norm of $\bm{\phi}_l(r)$ was less than predefined threshold,  set to $10^{-6}$ in our case. The final site densities can be determined as 
\begin{equation}
\bm{\rho}({\bf{r}})/\rho_0= \bm{\bar c}({\bf{r}})-\beta [\bm{S}_m \ast\bm{\phi}_l]({\bf{r}}) + 1
\end{equation}

Molecular dynamics calculations were based on the AMBER package \cite{Amber14} with
force field parameters are given in Table~1. All the simulations have been performed at $T=300$K. For N$_2$ simulations system was enclosed into 90 \text{\AA} cubic box containing 16050 solvent molecules, and for HCl simulation - 55 \text{\AA} cubic box containing 388 solvent molecules.

% We have employed  the smallest possible value of $\epsilon$ for the auxiliary sites, since the zero value of this parameter can not be used in the MD simulations.
% The standard combination
% rules, i.e. $\sigma_{ij}=\sigma_i/2+\sigma_j/2$, $\epsilon_{ij}=\sqrt{\epsilon_i\epsilon_j}$, are employed for the LJ 
% terms.   \textcolor{blue}{The system has been simulated ?? ns,
% the cut-off of potentials was ?? \AA.} 

\begin{table}[h!]
Table 1. Force field parameters of solute and solvent sites 
\begin{tabular}{|p{4cm}|p{3cm}|p{3cm}|p{3cm}|}
\hline\hline
\diagbox{sites}{parameters}     & $\epsilon$ (kcal/mol)&$\sigma$ (\AA)& $q$(e)   \\  \hline
\multicolumn{4}{|c|}{\textit{$N_2$ solvent  ($\rho_0=0.022\AA^{-3}$ \quad $l=1.135\AA$)}}  \\
N                               & 0.17                 & 3.25         & 0          \\                        
\hline
\multicolumn{4}{|c|}{\textit{$HCl$ solvent with auxiliary site  ($\rho_0=0.0234\AA^{-3}$ \quad $l=1.3\AA)$ }}                       
\\ 
H              & 0.0397     &    $1.8*10^{-4}$      & 0    \\ 
Cl             & 0.5138              & 3.35                & 0     \\ 
\hline 
\multicolumn{4}{|c|}{\textit{polar $HCl$ solvent   ($\rho_0=0.0234\AA^{-3}$ \quad $l=1.3\AA$) }}                       
\\ 
H              & 0 (0.0397)$^{a}$    & $1.8\cdot 10^{-4}$  & 0.2    \\ 
Cl             & 0.5138              & 3.35                & -0.2     \\ 
\hline 
 \multicolumn{4}{|c|}{\textit{Solute Parameters}}                    \\ \hline
 
N$_s$             & 0.17                & 1.625                & 0       \\ 
N$_l$            & 0.17                & 6.52                 & 0      \\ 
S            & 0.0397              & 0.00                 & 0       \\ 
L            & 0.5138              & 3.35                 & 0       \\ 
$Li^{+}$      & 0.0165              & 1.56                & 1.0       \\
$Cl^{-}$      & 0.5138              & 3.35                 & -0.2\\
\hline
\end{tabular}
\end{table}

\section{Example Applications}
\subsection{Liquid nitrogen}
\begin{figure}[t!]
\includegraphics[width=7in]{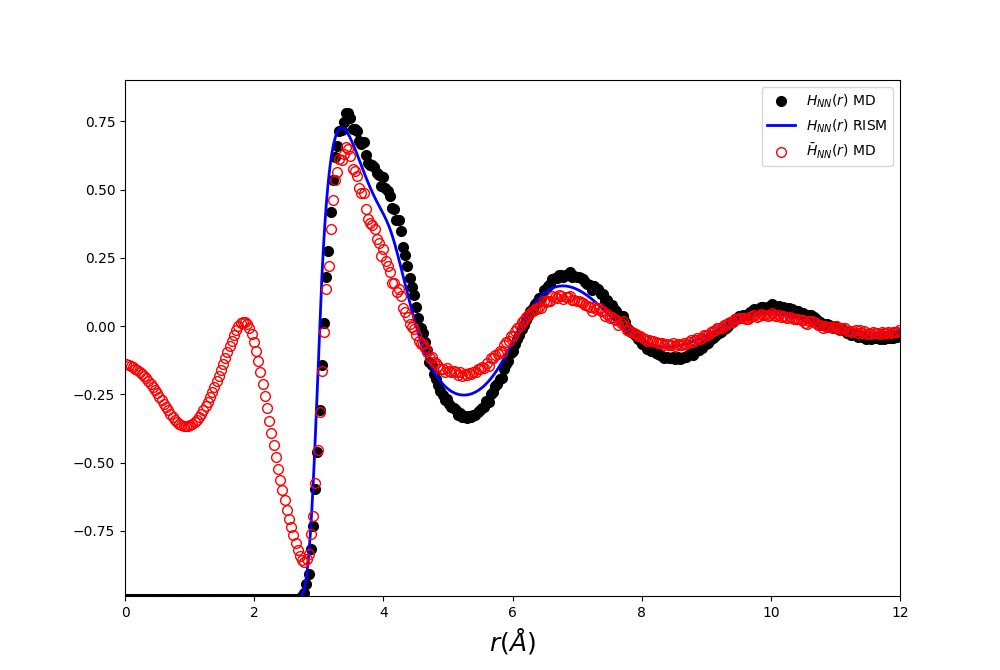}
\caption{Comparison of bare ($H_{NN}$)   and  renormalized ($\bar H_{NN}$) correlation correlation functions  for liquid nitrogen at $T=300 $K obtained from MD simulations and RISM calculations.}
 \label{fig1}
\end{figure}

Our first application involved $N_2$ liquid, with thermodynamic input data and force field parameters provided in Table~1. We note that, given identical atomic sites in this case, the matrix elements of bare correlation function become equal to each other, $H_{ij}(r) = H_{NN}(r)$. The resulting renormalized correlation function in $k$-space given by
\begin{equation}
\bar H_{NN}(r)=\frac{2}{\pi^2} \int \frac{H_{NN}(k)}{[1+d(k)]^2}\frac{\sin(kr)}{r}kdk
\end{equation}      
The comparison between bare and renormalized correlation functions is provided on Fig. 1. We observe that the main difference between the two functions is inside the nitrogen core region. The renormalized correlation function in this region acquires an additional oscillations related to additional intra-molecular correlations introduced through the renormalization procedure. 

\begin{figure}[t!]
\includegraphics[width=7.1in]{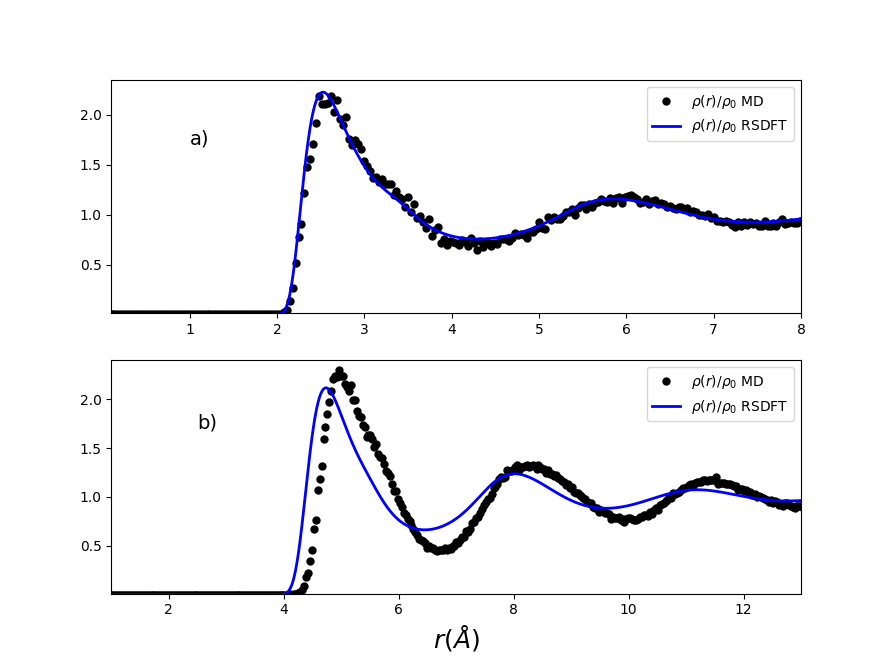}
\caption{Dimensionless site densities $\rho(r)/\rho_0$ for solutes $N_s$ (a) and $N_l$(b) solvated in liquid $N_2$ at $T=300$K which are obtained by the RSDFT (lines) and from the MD simulations (circles).
}
 \label{fig2}
\end{figure}

Using the above RCF, we have performed density profile calculations of inhomogeneous N$_2$ liquid for two different size solutes, N$_s$ and N$_l$. Results of these calculations together with the data obtained from the MD simulations are shown on Fig. 2. Overall, we observe a very good agreement between density profiles calculated with the MD and the RSDFT methods. For small solute (N$_s$) the two data sets are essentially the same. For larger solute (N$_l$), the RSDFT density profile is shifted slightly towards the solute. The reason for these differences likely comes from HNC approximation, which becomes less accurate for larger perturbations introduced by N$_l$ solute.

\subsection{Auxiliary site model for diatomic liquids}

In the auxiliary site model of diatomic liquids only one atomic site is physical, i.e. participating in all the interactions. The other site is auxiliary and is subject to only intra-molecular interactions, which come from its chemical (in this case rigid) bond to the physical site. The interest to this model is twofold. First, it provides the simplest way to include molecular features into simple liquid primitive models. Second, this model, as we demonstrate below,  can be rigorously mapped to a simple liquid system, and hence all the molecular
properties of such liquid can be derived from the data on the behaviour of the corresponding simple liquid. 

As a particular example of such system, we consider HCl-like diatomic liquid, where H and Cl represent auxiliary and physical sites correspondingly (see Table~1 for parameters). Per our procedure, we again calculate renormalized correlation functions from MD simulations of homogeneous systems. The results of these calculations are presented on Fig. 3. We observe that the hydrogen-hydrogen and hydrogen-chlorine renormalized correlation functions ($\bar{H}_{HH}$ and $\bar{H}_{HCl}$)  are essentially zero. Qualitatively,  this result can be understood by  recalling that our renormalization procedure essentially reduces  molecular liquid system to an effective simple fluid mixture consisting of atomic sites (see Section \ref{sec:renormalization}). Given the auxiliary nature of H-site, it is thus not surprising that it does not partake in the latter, and that our effective system contains only Cl sites. To put this on more quantitative ground, we note that the following relationships hold true for our auxiliary site model
\begin{eqnarray}
{H}_{HH}(k)=d^2(k)H_{ClCl}(k) \qquad {H}_{HCl}(k)=d(k)H_{ClCl}(k) 
\label{Mo3}
\end{eqnarray}

Substituting this into (\ref{H_bar}) we obtain the expected result
\begin{align}
    \bar{H}_{HH}(k) &= \bar{H}_{HCl}(k) = 0 
    \nonumber
    \\
    \bar{H}_{ClCl}(k) &= H_{ClCl}(k)
    \label{H_bar_HCl}
\end{align}

\begin{figure}[t!]
\includegraphics[width=7in]{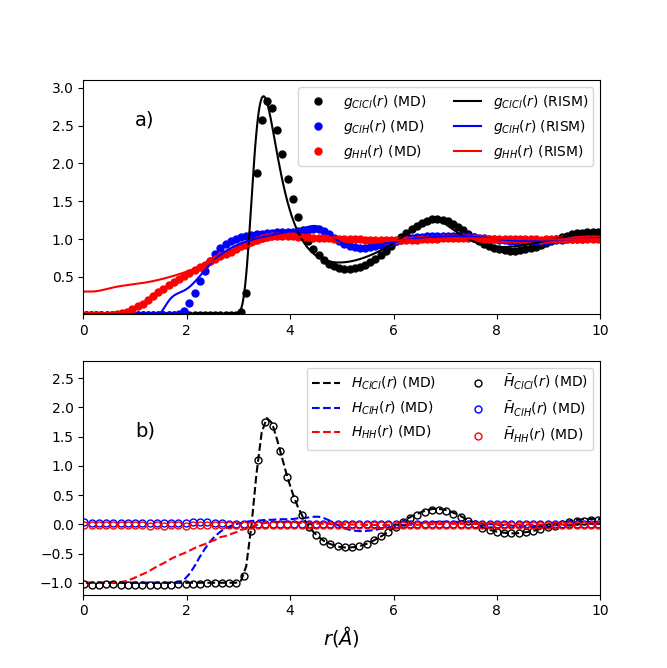}
\caption{Site-site radial distribution functions $g_{ij}(r)$, bare  correlation functions $H_{ij}(r)$, and  renormalized correlation functions $\bar H_{ij}(r)$ 
for nonpolar $HCl$  at $T=300$K  obtained from MD simulations (symbols and dashed lines) and by the 1D RISM-HNC equations (solid lines).}
 \label{fig3}
\end{figure}

\begin{figure}[t!]
\includegraphics[width=7.2in]{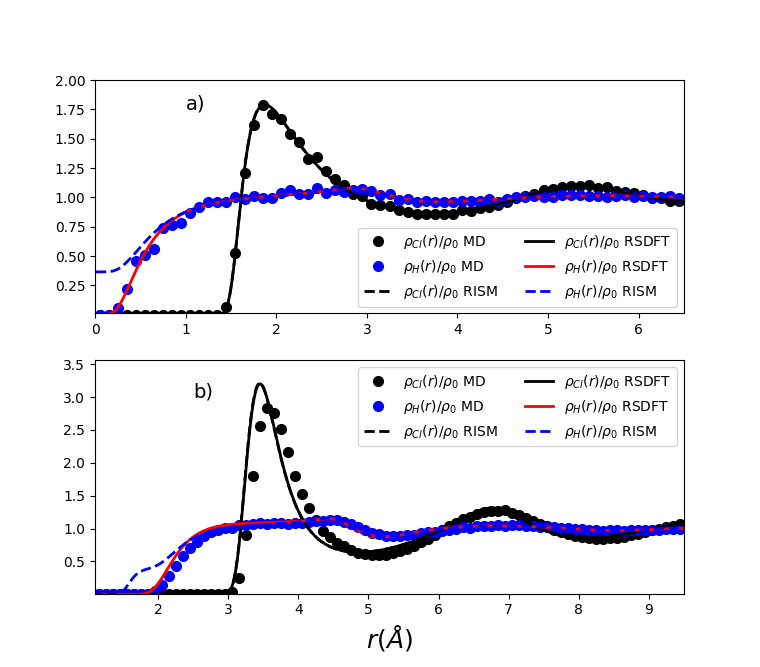}
\caption{Dimensionless site densities $\rho(r)/\rho_0$  for solutes $S1$ (a) and $L2$ (b) solvated in nonpolar $HCl$  at $T=300^0$K, obtained by the RSDFT (solid lines), RISM (dashed lines) and from the MD simulations (symbols).}
 \label{fig4}
\end{figure}

Based on the above RCF, we evaluate the site density profiles for two spherical solutes $S$ and $L$ (force field parameters are given in Table~1). The results are presented in Fig.~4 together along with data obtained from the MD simulations and also RISM theory. 
We observe that RSDFT results are in excellent agreement with MD results. The
RISM on other hand shows sizeable deviations for small distances due to improper description of the correlation hole effect.\cite{Chuev2020}

One important property of the auxiliary site model of inhomogeneous diatomic liquids is that the densities of the auxiliary and physical sites are related as\cite{Chuev2020}
\begin{eqnarray}
{\rho}_{H}(k)=d(k)\rho_{Cl}(k) 
\label{Mo}
\end{eqnarray}
\begin{figure}[t!]
\includegraphics[width=7.2in]{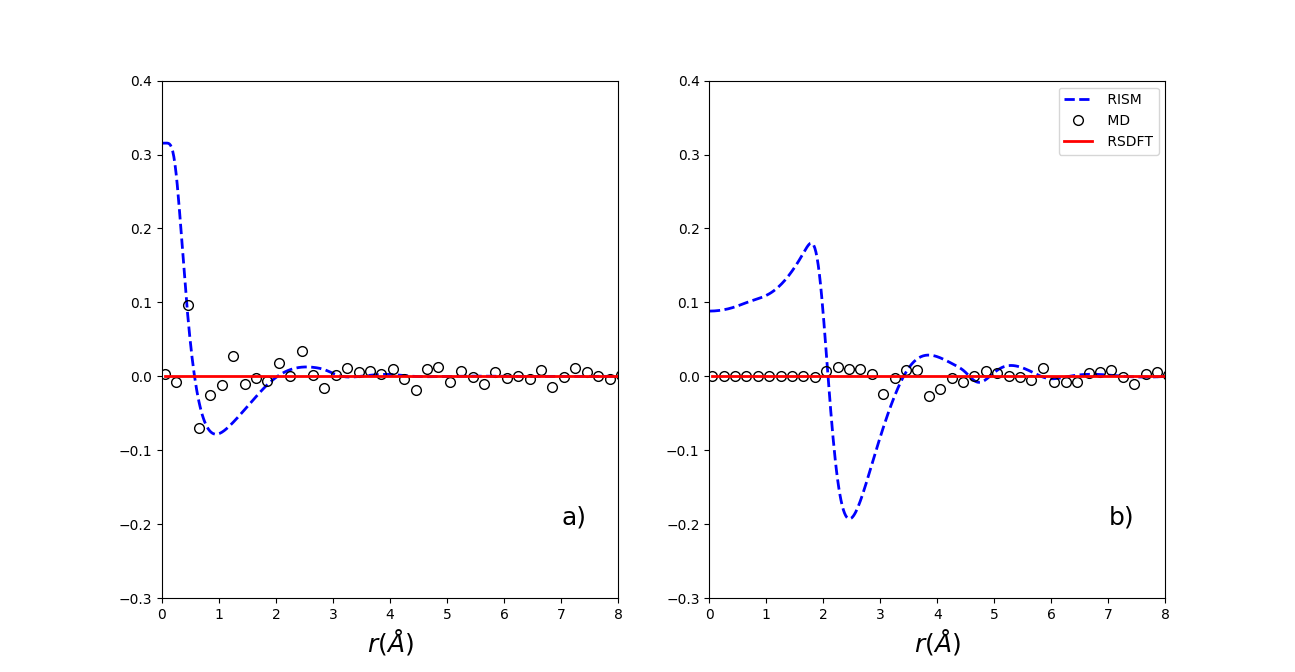}
\caption{Differences between the calculated and projected  
(\ref{Mo} )  hydrogen site densities for solutes $S$ (a) and $L$ (b) 
solvated in HCl molecular liquid at $T=300$K. The dashed lines correspond to the RISM,
solid lines to the RSDFT, whereas the symbols to the data obtained from the MD simulations.}
 \label{fig5}
\end{figure}
The above relationship, along with our earlier results for renormalized correlation function, suggests that analysis of the inhomogenous molecular liquid with auxiliary sites can be reduced to that of  
simple liquid consisting involving only physical site. To demonstrate this, we have performed standard integral equation theory (IET) calculations for inhomogeneous simple liquid consisting only of Cl atoms, under the same density and temperature as our molecular liquid. The resulting density of the H site, computed through Eq.(\ref{Mo}), was then compared with MD, RSDFT, and RISM data (see Fig.~5). We observe that this simple liquid based procedure is indeed in an excellent agreement with both MD and RSDFT predictions. The deviations from RISM predictions at small distances stem from the inaccuracies of latter as have been discussed above.

\subsection{Polar diatomic liquid}

\begin{figure}[t!]
\includegraphics[width=7in]{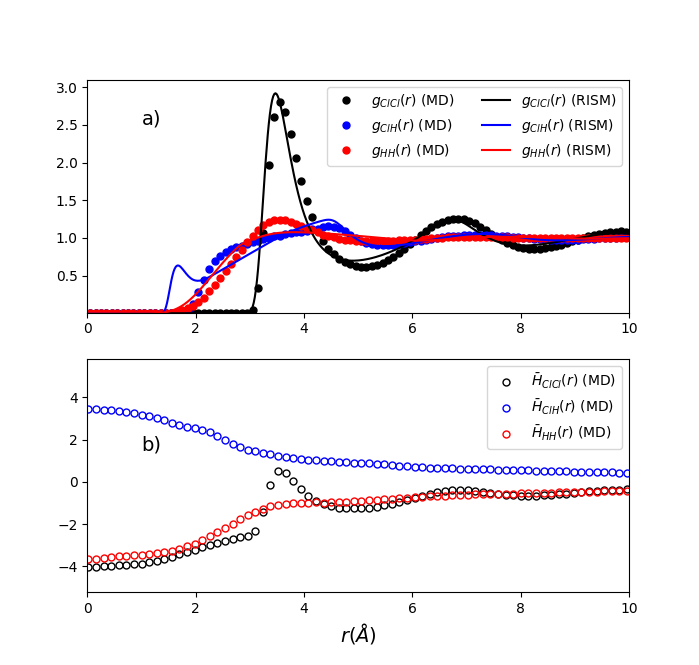}
\caption{Site-site radial distribution functions $g(r)$ and renormalized
correlation functions $\bar H(r)$ 
for polar $HCl$  at $T=300$K  obtained from MD simulations (symbols) and by the 1D RISM-HNC equations (lines).}
 \label{fig6}
\end{figure}

As our final application, we consider a polar 
diatomic liquid, which is obtained by adding charges to auxiliary site model considered in the previous section. These types of systems are particularly difficult to treat with existing RISM methods, due to delicate balance between Coulomb forces on, now charged, auxiliary H site and its chemical bond to physical
site. The problem is illustrated on Fig. 6, which compares of RDF's 
obtained by MD and RISM approaches. While RISM does a reasonable job in describing $g_{ClCl}$, the accuracy of distributions involving H site are markedly worse. A particular worrisome feature is the appearance of artificial peak in
$g_{ClH}$ RDF at ~1.8 {\AA}. Aside the fact that this is not observed in MD simulation, such peak would result in Cl-Cl distance of 3.1 {\AA}, which would be highly improbable according to the same RISM calculation.

\begin{figure}[t!]
\includegraphics[width=7.1in]{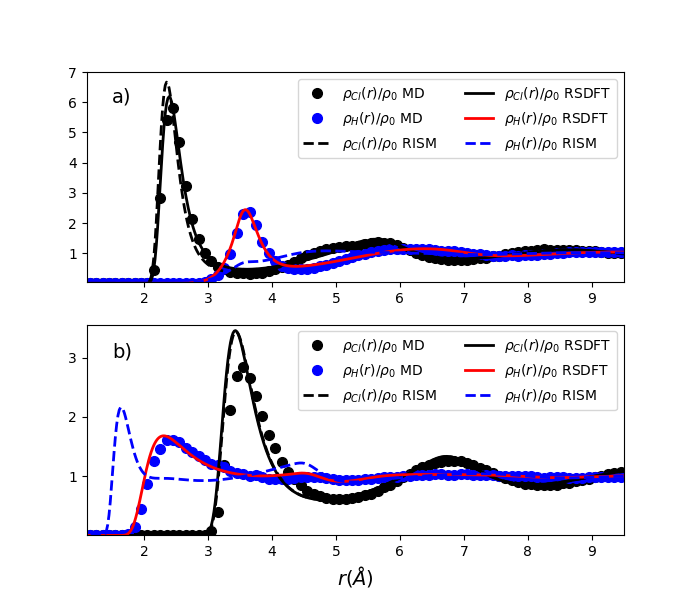}
\caption{Site density distributions for polar HCl liquid in the presence  solutes of cation $Li^+$ (a) and anion $Cl^-$ (b) solutes obtained by the RSDFT (solid lines), RISM (dashed lines) and MD simulations (circles).}
 \label{fig4}
\end{figure}

As indicated on Fig. 6, the renormalized correlation functions  $\mathbf{\bar{H}}$ for polar HCl liquid no longer resemble those found in the auxiliary model. None of the components vanish and all exhibit much slow decay consistent with Coulomb interactions. To test performance of RSDFT calculations we considered two types of charged solutes - small cation (Li$^+$) and large anion (Cl$^-$). The resulting site densities are shown on Fig. 7.
As in all the cases considered earlier, we have an excellent agreement between RSDFT and MD data. For both cation and anion solutes, RSDFT corrects the existing deficiencies of RISM approach. This improvement comes simply as a result of consistent treatment of intra-molecular correlation effects. This should be contrasted with a posteriori corrections utilized in RISM applications, such as repulsive bridge correction  \cite{Du2000,Kovalenko2000b}. While the latter may mimic negative correlation hole effect \cite{Chuev2020} for anion case, it will certainly make matters worse for cation case, where correlation hole effect is of the opposite sign.

\section{Discussion and Conclusions}

Dual space formulation introduced in this work builds upon conventional SDFT approach, incorporating independent field variable into site-density representation. Similar ideas have been used in the past in simple liquid theory.\cite{Caillol_2002} In our case this gives as a flexibility to evaluate
intra-molecular interactions in more appropriate field based representation, which significantly
simplifies analysis of molecular liquid system. As a result of this, the SDFT analysis of the molecular liquid problem reduces to the solution of system of self-consistent equations, similar to Kohn-Sham equations in electronic structure DFT.

Focusing on a particular case of density-based homogeneous reference expansion of inter-molecular interactions, we demonstrate that site density description at the long range (low-k regime) scale becomes ill-conditioned or quasi-degenerate, ultimately resulting in divergences. In hindsight, such behaviour was to be expected, reflecting a simple fact that as we approach macroscopic scale, the atomistic details become difficult to resolve. To mitigate this issue, we define new collective variables, renormalized response functions, which are natural generalizations of direct correlation functions used in simple fluids. Transformation to these new variables can be viewed as a renormalization procedure that makes our molecular liquid system to look more like a  simple fluid mixture. The interaction in the latter are driven by dressed or renormalized inter-molecular correlation functions (RCF's),  providing, in essence, a mechanism to propagate details of interactions at the molecular scale to long range inter-molecular scale. 

The developed  renormalization  procedure delegates the problem of quasi-degeneracy of site-density basis  to the construction  of  proper  RCFs. With a clear  meaning as effective interactions, the RCF's present a physically appealing way to bypass numerical problems. The particular procedure utilized in this work regularizes RCF's by changing their asymptotic behavior at low-k wave limit. The latter does not affect the quality of solution when a regularization parameter is small enough. 

The resulting approach, which we refer to as renormalized SDFT or in short RSDFT, was applied to several types of diatomic liquids, probing various potential scenarios (identical atomic sites, auxiliary atom type, and charged species). We demonstrate that RSDFT,  being as computationally efficient as the 3D-RISM approach, yields substantially better results especially in the case of auxiliary sites and hydrogen bonded solutes.

In work we have considered only 2nd order terms in the density expansion 
of inter-molecular correlation functional. Given that such expansion is only applied to inter-molecular part of the interaction potential, and not to the entirety of it as done in RISM methods, it should captures major portion of inter-molecular correlation effects. This is clearly evidenced in our application results as well. Still the inclusion of higher order bridge corrections may prove to be necessary and can be investigated within our proposed framework.

The diatomic molecular liquids considered in this work provides just the first initial 
test for the performance of our approach. While this may already prove useful 
in extending existing primitive models of molecular solutions, ability to treat general poly-atomic systems would be important next step. Our framework has no  limitations in that respect, yet the computations of correlation hole involving angular degrees of freedom may require some care.  

\acknowledgments
We would like to thank Chris Mundy, Greg Schenter and Shawn Kathmann for helpful discussions. M.V. was supported by the U.S. Department of Energy, Office of Science, Office of Basic Energy Sciences, Division of Chemical Sciences, Geosciences, and Biosciences. PNNL is operated by Battelle Memorial Institute for the United States Department of Energy under DOE Contract Number DE-AC05-76RL1830. The reported study was also partially funded by the RFBR (G.N.Ch. and M.V.F.).

\appendix
\section{Dual space functional}
\label{app:sec_M}
Let us consider density functional corresponding to the molecular gas system. As an
alternative to standard Legendre expression, it can also be written as a more general Legendre-Fenchel
transformation\cite{Rockafellar1970} of the molecular field functional (\ref{Wm_general}) as
\begin{align}
    \Gamma_{m}[\bm{\rho}] = \sup_{\bm{J}}\{
    \Omega_{m}\left[\bm{J}\right]
    -\bm{J} \cdot \bm{\rho}
    \}
    \label{app:gamma_m_as_sup}
\end{align}
where $\bm{J}$ and
$\bm{\rho}$ are now considered to be an independent variables. Combining 
(\ref{app:gamma_m_as_sup}) with (\ref{upsilon}) we arrive at the following expression
for the density functional of the molecular liquid system 
\begin{align}
\Gamma[\bm{\rho}]=
    \sup_{J}\{
    \mathcal{M}[\bm{J},\bm{\rho}]
    \}
    \label{app:gamma_as_sup}
\end{align}
where
\begin{align}
    \mathcal{M}[\bm{J},\bm{\rho}] = 
    \Omega_{m}\left[\bm{J}\right]
    -\bm{J} \cdot \bm{\rho}
    +\Upsilon_l[\bm{\rho}] 
\end{align}
Equation (\ref{app:gamma_as_sup}) combined with the
the variational principle (\ref{var}) leads to the desired result that solution 
to our molecular liquid system can be found by variation of dual field functional
with respect to field $\bm{J}$ and density $\bm{\rho}$:
\begin{equation}
\frac{\delta \mathcal{M}[\bm{\xj},\bm{\rho}]}{\delta \bm{\xj}({\bf{r}})}=0
\qquad
\frac{\delta \mathcal{M}[\bm{\xj},\bm{\rho}]}{\delta \bm{\rho}({\bf{r}})}=0
\end{equation}

\bibliography{marat}
\end{document}